\documentclass[12pt]{article}
\usepackage{amsmath}
\usepackage{amsfonts}
\usepackage{amssymb}
\usepackage{hyperref}

\numberwithin{equation}{section}

\newcommand{\ns}{\normalsize} 
\newcommand{\ax}{\alpha}
\newcommand{\bx}{\beta}
\newcommand{\cx}{\gamma}
\newcommand{\dx}{\delta}

\newcommand{\nn}{\nonumber}
\newcommand{\Bs}{\mathcal{B}}

\begin{document}

\begin{titlepage}

\title{
    \vskip 2cm
   {\Large\bf F-theory compactifications on manifolds with $SU(3)$ structure}\\[1.5cm]}

\author{{\bf Andrei Micu} \\[1cm]
   {\it\ns Horia Hulubei National Institute of Physics and Nuclear
     Engineering -- IFIN-HH}\\  
   {\ns Str. Reactorului 30, P.O.~Box MG-6, M\u{a}gurele, 077125, jud
     Ilfov, Romania}\\
   {\tt\ns amicu@theory.nipne.ro}\\[1cm]
}
\date{}

\maketitle
\begin{abstract}
  In this paper we derive part of the low energy action corresponding to
  F-theory compactifications on specific eight manifolds with $SU(3)$
  structure. The setup we use can actually be reduced to
  compactification of six-dimensional supergravity coupled to tensor
  multiplets on a $T^2$ with duality twists. The resulting theory is a
  $N=2$ gauged supergravity coupled to vector-tensor multiplets.
\end{abstract}

\thispagestyle{empty}

\end{titlepage}

\section{Introduction}

Recently it was pointed out that in the presence of certain 
fluxes, the heterotic --type IIA duality in four dimensions requires that, 
on the type IIA side, M-theory has to be considered instead. The fluxes which 
are responsible for this behaviour are ordinary fluxes for the heterotic 
gauge fields \cite{ABLM}. The full duality picture is heterotic string
compactified on  $K3 \times T^2$ with duality twists is
the same as M-theory compactified on seven-dimensional manifolds with
$SU(3)$ structure which are obtained by fibering Calabi--Yau manifolds
over a circle \cite{AM1}. 

It turns out that the heterotic picture can be further generalised by
allowing twists in the full 4d-duality group
\cite{RES}. This construction gives what is known under the name of
R-fluxes \cite{STW}. It has been conjectured that the dual of this
setup can be found 
in F-theory compactifications on eight-dimensional manifolds with
$SU(3)$ structure obtained by fibering a Calabi--Yau manifold over a
$T^2$ much in the same way as it was done in the M-theory case
\cite{AM1}. Motivated by this, we study the tensor multiplet sector of
such F-theory compactifications. This leads to $N=2$ supergravity theories
in four dimensions coupled to vector-tensor multiplets.

\section{General setup}
\label{sec:gensetup}

We are interested in F-theory compactifications on eight-dimensional
manifolds with $SU(3)$ structure obtained by fibering a Calabi--Yau
manifold over a two torus, $T^2$. The fibration is done such that the
two-forms on the Calabi--Yau manifold satisfy
\begin{equation}
  \label{do}
  d \omega_\alpha = - M_i^\beta{}_\alpha \omega_\beta \wedge dz^i \; .
\end{equation}
where $i,j = 1,2$ denote the torus directions, while $\omega_\alpha$
denote the harmonic two-forms on the Calabi--Yau manifold and the
matrices $M_1$ and $M_2$ are constant commuting matrices which are in
the algebra of the symmetry group on the space of two-forms.

Since there is no low energy effective action description for
F-theory, a direct compactification is not possible and we have to
rely on other methods. In particular for the case above, the fibration
can be effectively realised by splitting the compactification into 
a compactification on a Calabi--Yau
three-fold followed by a Scherk-Schwarz compactification \cite{SS} on
the torus. 
After the first step, the the six-dimensional fields which come from an
expansion in the forms $\omega_\alpha$ which satisfy \eqref{do} would
have a non-trivial dependence on the torus coordinates, which is why
one has to consider a Scherk--Schwarz compactification in
order to obtain the correct result.

Let us specify more the compactification Ansatz. We consider
throughout that the Calabi--Yau three-fold is elliptically fibered
with four-dimensional base $\mathcal{B}$.
The two-forms may have two origins: two forms
which come from the base of the fibration and two-forms which come from
resolving the singularities of the fibration. In the following we shall
concentrate only on the first type of two-forms, namely the ones which
already exist on the base of the fibration. It is known that in F-theory
compactifications on Calabi--Yau 3-folds these forms lead in six
dimensions to antisymmetric tensor fields. It is precisely this
tensor-field/ tensor-multiplet sector that will be of interest for us
in the following. 

If we denote the number of $(1,1)$ forms on $\Bs$ by $h^{1,1}(\Bs)$ then,
$T$, the number of tensor multiplets is given by $ T= h^{1,1}(\Bs) -1$.
Note that supersymmetry requires that $h^{2,0}(\Bs)=0$, and therefore,
all the two forms of interest -- and in particular the forms in
\eqref{do} -- are the $(1,1)$ forms on $\Bs$. On such a
four-dimensional space there is precisely one self-dual $(1,1)$ form
(the K\"ahler form) and $T$ anti-self-dual $(1,1)$ forms. This implies
that the inner product on the space of two-forms posseses a $SO(1,T)$
symmetry. This symmetry is nothing but the symmetry found in
\cite{romans} on the space of tensor-fields in six-dimensional
$N=1$ supergravity coupled to $T$ tensor multiplets. Therefore, we
choose the twist matrices $M_1$ and $M_2$ to be generators of
$SO(1,T)$.

Let us summarize. We have just argued that F-theory compactifications
on eight-dimensional manifolds obtained by fibering a Calabi--Yau
manifold over a torus as described in \eqref{do} can be effectively
modeled by considering six dimensional compactifications of F-theory
followed by a compactification on a torus with $SO(1,T)$ duality twists.
In particular we shall be interested in tensor-multiplet sector of the
six-dimensional theory.

\section{Compactification with duality twists}

\subsection{The six-dimensional theory}

Let us start by describing the content of the theory in six
dimensions. A similar description of the theory appeared recently in
\cite{thomas}. We are interested in six-dimensional minimal supergravity
coupled to $T$ tensor multiplets. We suppose throughout that the
number of hypermultiplets is such that the gravitational anomalies are
canceled. The supergravity multiplet contains as bosonic degrees of
freedom the graviton $g_{\mu\nu}$ and an antisymmetric tensor field
with self-dual field strength. Each of the tensor multiplets contain
as bosonic degrees of freedom one antisymmetric tensor field with
anti-self-dual field strength and one scalar field. The
(anti-)self-duality of these tensor fields can also be seen from the
F-theory/type IIB compactification. Recall that type IIB string
features in ten dimensions a RR four-form potential, $C_4$, with
self-dual field strength. When expanded in the $(1,1)$ harmonic forms
on the base $\Bs$ of the Calabi--Yau three-fold this precisely yields one
tensor field with self-dual field strength and $h^{1,1}(\Bs)-1 \equiv
T$ tensor fields with anti-self-dual field strengths. 

Let us denote all the tensor fields generically by $B^\alpha, \alpha =
1, \ldots , T+1$ and the K\"ahler moduli corresponding to deformations
of the base by $v^\alpha$. These fields appear from the expansion of
the RR four-form $C_4$ and of the K\"ahler form $J$ in a basis of
$(1,1)$ harmonic forms on the base $\mathcal{B}$.
\begin{equation}
  \label{CJexp}
  C_4 = \ldots + B^\alpha \omega_\alpha + \ldots \; ; \qquad J=
  v^\alpha \omega_\alpha \; .
\end{equation}
Note that we work with a basis of $(1,1)$ forms in which the (anti)
self-duality is not manifest.
Let us define the intersection numbers on $\mathcal{B}$ by
\begin{equation}
  \label{intB}
  \rho_{\alpha \beta} = \int_B \omega_\alpha \wedge \omega_\beta \; .
\end{equation}
The matrix $\rho$ has $(1,T)$ signature and is the matrix which is
used to raise and lower $SO(1,T)$ indices. 
The volume of the base which is defined as
\begin{equation}
  \label{volB}
  \mathcal{V} = \tfrac12 \int_B J \wedge J = \tfrac12 \rho_{\alpha\beta}
  v^\alpha v^\beta \; ,
\end{equation}
is part of a hypermultiplet. Therefore in order to correctly describe
the number of $T$ scalar degrees of freedom by $T+1$ variables $v^\ax$
we shall work at constant volume, $\mathcal{V}=1$.

It has been known from
\cite{MS,romans} that these theories admit a manifestly Lorenz invariant
Lagrangean description only in the case $T=1$. For an arbitrary number
of tensor-multiplets -- and here we want to keep this number arbitrary
-- the self-duality conditions make it impossible to derive the theory
from an action principle. However, since we are only interested in the
four-dimensional compactified theory, we shall addopt a strategy,
which was used in type IIB compactifications \cite{dallAgata}, which
will allow us to circumvent the above problem. The idea is to write
down an action for tensor fields whose field strengths are not
constrained by any self-duality condition. In this way we double the
number of degrees of freedom described by the tensor fields. After the
compactification to four dimensions the additional degrees of freedom
manifest themselves as independent fields which are Poincare dual to
the normal degrees of freedom which we would have expected from the
compactification. By adding suitable Lagrange multiplier terms to the
action we can impose the four-dimensional version of the self-duality
conditions as the equations of motion for the additional degrees of
freedom in the theory. Eliminating at this step these degrees of
freedom from their equations of motion we obtain the theory we were
searching in the first place.

Therefore we consider the following starting six-dimensional
action\footnote{Hats $\hat{ } $ are used in order to distinguish
  six-dimensional fields from their four-dimensional descendants.}
\begin{equation}
  \label{6dact}
  S = - \tfrac12 \int \left( R + \tfrac12 g_{\ax\bx} \hat H^\ax \wedge * \hat
    H^\bx  + \left. g_{\ax \bx} d \hat v^\ax \wedge *d \hat v^\bx
    \right|_{\mathcal{V}=1} \right) \; , \quad  \ax, \bx = 1, \ldots,
  T+1 \; ,  
\end{equation}
where $\hat H^\ax$ denotes the field strength for the tensor fields
which is given by
\begin{equation}
  \hat H^\ax = d \hat B^\ax \; .
\end{equation}
The metric $g_{\ax \bx}$ can be seen as coming from the F-theory/type
IIB compactification as\footnote{Up to factors of $\mathcal{V}$ which
  are irrelevant as we set $\mathcal{V}=1$.}
\begin{equation}
  \label{g}
  g_{\ax \bx} = \int_\mathcal{B} \omega_\ax \wedge * \omega_\bx \; ,
\end{equation}
and has a $SO(1,T)$ isometry group.  In order to have the correct
theory we have to impose the self-duality conditions
\begin{equation}
  \label{SD}
  * \hat H^\alpha = \rho^{\alpha \beta} g_{\beta \gamma} \hat H^\gamma \; ,
\end{equation}
by hand as they can not be derived from the action \eqref{6dact}.
This relation is self-consistent precisely due to the $SO(1,T)$
symmetry  which ensures that
\begin{equation}
  g^{-1 \ \ax \bx } = \rho^{\ax \cx} g_{\cx \dx} \rho^{\dx \bx} \; .
\end{equation}
The above data specify the six-dimensional action. We shall use this
formulation in the next section in order to perform a compactification
on a torus with duality twists.

\subsection{Scherk-Schwarz compactification to four dimensions}

In this section we perform the Scherk--Schwarz toroidal
compactification of the six-dimensional theory presented before. Let
us start by describing the degrees of freedom we expect in the
four-dimensional theory. From the gravity sector there will be two
Kaluza--Klein vector fields $V^{1,2}$ and three torus moduli which we
shall take as the three independent components of the metric on the
torus $G^{11}$, $G^{12}$ and $G^{22}$. One of the vector fields will
be the graviphoton, the scalar superpartner of the graviton in four
dimensions, while the other vector field together with two of the
torus moduli will become the bosonic components of a vector
multiplet. 

From the tensor fields compactified on the torus we expect the
following degrees of freedom
\begin{equation}
  \label{Bexp}
  \hat B^\alpha = B^\alpha + A_i^\alpha \wedge dz^i + b^\alpha dz^1
  \wedge dz^2 \; .
\end{equation}
Due to the (anti)sef-duality condition which the corresponding
field-strengths satisfy the number of degrees of freedom is only half
of the ones above. In particular we expect one vector field and either
a scalar field or a tensor field. In case we keep the scalar we will
end up with a true vector multiplet while if we keep the tensor we
will have a vector-tensor multiplet. The additional scalars in these
multiplets are given by the remaining torus modulus above and the
scalars which already exist in six dimensions as superpartners of the
tensor fields. Altogether we will end up with a number of $T+2$ vector
plus vector-tensor multiplets. 

Let us now see how the compactification proceeds. As
explained in the previous section, the part of the theory we are
interested in has a $SO(1,T)$ duality symmetry. We shall use this
symmetry in order to perform the compactification with duality
twists. 
In particular we are interested in the following dependencies on the
internal coordinates of the torus
\begin{equation}
  \label{Btw}
  \begin{aligned}
  \partial_i \hat B^\alpha &  = M_i^\ax{}_\bx \hat B^\bx \; , \\
  \partial_i \hat v^\alpha & = M_i^\ax{}_\bx \hat v^\bx \; .
  \end{aligned}
\end{equation}
The sign difference compared to \eqref{do} comes from the fact that we
are addopting the passive rather than the active picture for
the symmetry transformations. Note that the $SO(1,T)$ transformation
of the volume of the base $\Bs$ is given by
\begin{equation}
  \delta \mathcal{V} = \rho_{\ax \bx} M_i^\ax{}_\cx \hat v^\cx \hat v^\bx +
  \rho_{\ax \bx} \hat v^\ax M_i^\bx{}_\cx \hat v^\cx \; ,
\end{equation}
which vanishes because the generators $(M_i)_{\alpha \beta} =
\rho_{\alpha\gamma} M_i^\gamma{}_\beta$ are antisymmetric in the
indices $\alpha$ and $\beta$, ie
\begin{equation}
  \label{Mas}
  \rho_{\alpha\gamma} M_i^\gamma{}_\beta + \rho_{\beta\gamma}
  M_i^\gamma{}_\alpha =0 \; .
\end{equation}
Note that this is consistent with the fact that the
volume of the base is part of a hypermultiplet.

Let us consider the standard metric for the compactification on $T^2$
\begin{equation}
  \label{metric}
  ds^2 = g_{\mu\nu} dx^\mu dx^\nu + G_{ij} (dz^i - V^i) (dz^j - V^j)
  \; ,
\end{equation}
where $g_{\mu\nu}$ is the metric on the four-dimensional space,
$G_{ij}$ is the metric on the torus and by $V^i, \ i =1,2$ we denoted
the Kaluza--Klein vector fields which come from the torus
compactification. 

The field strengths $\hat H^\alpha$ which come from the expansion
\eqref{Bexp} read
\begin{equation}
  \label{Hexp}
  \hat H^\alpha = d B^\alpha + \left( d A_i^\alpha + M_i^\alpha{}_\beta
  B^\beta \right) \wedge dz^i + \left( d b^\alpha + M_2^\alpha{}_\beta
  A_1^\beta - M_1^\alpha{}_\beta A_2^\beta \right) dz^1 \wedge dz^2 \; ,
\end{equation}
while for the scalar fields $v^\ax$ we find
\begin{equation}
  \label{dbexp}
  d \hat v^\alpha = d v^\alpha + M_i^\ax{}_\bx v^\bx dz^i \; .
\end{equation}

Note that in toroidal compactifications the basis for field expansions
$dz^i$ is not invariant under rezidual diffeomorphism transformations
-- which induce four-dimensional gauge transformations -- and
therefore the fields which result 
from this expansion will have non-standard transformation
properties. This has also a rather technical consequence. Since Hodge
$*$ operator in \eqref{6dact} is taken with respect to the metric
\eqref{metric} which is non-diagonal, it will introduce crossed
terms between the four-dimensional space and the torus. A
factorization can still be achieved if we use for the 
expansion of the fields involved in \eqref{6dact} the gauge invariant
basis 
\begin{equation}
  \label{eta}
  \eta^i = dz^i - V^i \; .
\end{equation}
This is precisely the basis which should be used in order to obtain
fields with correct gauge transformations. Rewriting the field
strengths \eqref{Hexp} and \eqref{dbexp} in this basis we obtain
\begin{equation}
  \label{fsexp}
  \begin{aligned}
  \hat H^\alpha & = H^\alpha + F_i^\ax \wedge \eta^i + D b^\ax
  \eta^1 \wedge \eta^2 \; , \\
  d \hat v^\alpha & = D v^\alpha - M_i^\ax{}_\bx v^\bx \eta^i \; ,
  \end{aligned}
\end{equation}
where
\begin{eqnarray}
  \label{fsdef}
  H^\alpha & = & dB^\ax + F_1 \wedge V^1 + F_2 \wedge V^2 - D b^\ax \wedge V^1
  \wedge V^2 \; ; \nonumber \\
  F_1^\alpha & = & dA_1^\ax + M_1^\ax{}_\bx B^\bx - Db^\ax \wedge V^2
  \; ; \quad
  F_2^\alpha  =  dA_2^\ax + M_2^\ax{}_\bx B^\bx + Db^\ax \wedge V^1
  \; ; \nn \\
  Db^\ax & = & d b^\ax- M_1^\ax{}_\bx A_2^\bx + M_2^\ax{}_\bx A_1^\bx
  \; ; \quad
  Dv^\ax  =  d v^\ax + M_i^\ax{}_\bx v^\bx V^i \; . 
\end{eqnarray}

For the forms on the torus we use the following normalisation $
\int_{T^2} \eta^1 \wedge \eta^2 =1$, which implies
\begin{equation}
  \label{intt2}
  \int_{T^2} \eta^i \wedge * \eta^j = \sqrt G G^{ij} \; ,
  \quad \int_{T^2} * 1 = G \int_{T^2} \eta^1 \wedge \eta^2 \wedge
  *(\eta^1 \wedge \eta^2) = \sqrt G \; .
\end{equation}
Performing the integration over the torus the tensor field part in the action
\eqref{4dact} becomes
\begin{equation}
  \label{4dact}
  S_{T} = - \tfrac14 \int \sqrt G \left(g_{\ax \bx} H^\ax \wedge *H^\bx +
    g_{\ax\bx} G^{ij} F^\ax_i \wedge * F^\bx_j + \frac{1}{G} g_{\ax
      \bx} Db^\ax \wedge * D b^\bx \right) \; .
\end{equation}
To this we have to add the part of the action which descents from the
six-dimensional Ricci scalar  and from the kinetic term of the scalars
$v^\ax$
\begin{equation}
  \label{SR}
  S_R = -\tfrac12 \int \sqrt{G} \left(R + G_{ij} dV^i \wedge * dV^j +
    d G_{ij} \wedge * d G^{ij} + g_{\ax \bx} D v^\ax \wedge * D v^\bx
    + V  \right) \; , 
\end{equation}
The potential $V$ comes from the second term in the expansion of
$d \hat v^\ax$ in \eqref{fsexp} and is given by
\begin{equation}
  V = G^{ij} g_{\ax \bx} M_i^\ax{}_\dx M_j^\bx{}_\cx
  v^\dx v^\cx \; .
\end{equation}
The correct four-dimensional theory is obtained only after imposing
the self-duality conditions \eqref{SD}. Inserting the expansion
\eqref{fsexp} into \eqref{SD} we obtain their four-dimensional
analogues 
\begin{equation}
  \label{4dsd}
  \begin{aligned}
    \rho_{\ax \bx} Db^\bx & = \sqrt G g_{\ax \bx} * H^\bx \; ;\\
    \rho_{\ax\bx} *F^\bx_i & = \sqrt G \epsilon_{ij} G^{jk} g_{\ax \bx}
    F^\bx_k \; .
  \end{aligned}
\end{equation}
We see that these self-duality conditions identify a field with its
Poincare dual, as also explained at the beginning of this section
Evaluating the second relation above for explicit values of the
indices $i, j = 1,2$, we obtain 
\begin{equation}
  \label{Asd}
  \begin{aligned}
  \frac1{\sqrt G}* F^\ax_1 & = G^{21} \rho^{\ax \bx} g_{\bx \cx} F^\cx_1 + G^{2,2}
  \rho^{\ax \bx} g_{\bx \cx} F^\cx_2 \; , \\
  \frac1{\sqrt G} * F^\ax_2 & = - G^{11} \rho^{\ax \bx} g_{\bx \cx} F^\cx_1 - G^{1,2}
  \rho^{\ax \bx} g_{\bx \cx} F^\cx_2 \; ,    
  \end{aligned}
\end{equation}
which can be easily checked that are equivalent.

\subsection{Gauge transformations}

It will be instructive to collect the gauge transformations for the
fields in four dimensions. There are two types of symmetries that we
can find. First of all there is the gauge symmetry associated with the
tensor fields in six dimensions $\delta \hat B^\alpha = d
\hat \Lambda^\alpha$ for some gauge parameters $\hat \Lambda^\alpha$
which are one-forms. For these gauge parameters we have to consider a
dependence 
on the torus coordinates which similar to \eqref{Btw}. The easiest
way to see this is to consider the compactification on the full
eight-dimensional manifold described by \eqref{do}. Then the gauge
invariance above can be obtained by recalling the origin \eqref{CJexp}
of the tensor fields and using the gauge invariance of the four-form $C_4$
\begin{equation}
  \delta C_4 = d \Lambda_3 = d(\Lambda^\alpha \wedge \omega_\alpha +
  \lambda^\alpha_i \omega_\alpha \wedge dz^i) \; ,
\end{equation}
where $\Lambda^\ax $ is a 1-form while $\lambda^\ax_i$ are scalar functions.
Using this, equation \eqref{do} and the expansion \eqref{Bexp} we can
directly read off the transformations of the four-dimensional fields
\begin{eqnarray}
  \label{gaugetrsf1}
  \delta B^\alpha & = & d \Lambda^\alpha \; , \qquad \delta
  A^\alpha_i = - M_i^\alpha{}_\beta \Lambda^\beta. \; ; \\
  \label{gaugetrsf2}
  \delta A^\alpha_i & = & d \lambda^\alpha_i \; , \qquad \delta
  b^\alpha = M_1^\alpha{}_\beta \lambda^\beta_2 - M_2^\alpha{}_\beta
  \lambda^\beta_1 \; .
\end{eqnarray}
It is obvious that $Db^\alpha$ is invariant under the gauge
transformation \eqref{gaugetrsf2}, while invariance under
\eqref{gaugetrsf1} is guaranteed by the fact that the matrices $M_i$
commute. With this remark 
it is clear that the other field strengths, $ H^\alpha$ and
$ F^\alpha_{1,2}$ are also invariant under \eqref{gaugetrsf1} and
\eqref{gaugetrsf2}. 

The second gauge symmetry we discuss originates from the residual
diffeomorphism invariance on the torus. Under the infinitesimal
transformation $\delta z^i = \epsilon^i$, the KK gauge fields change
as $\delta V^i = d \epsilon^i$ leaving the one-forms $\eta^i$,
\eqref{eta}, invariant. The fields from \eqref{Bexp} however do not
have good transformation properties. In order to obtain fields whose
transformations do not involve derivatives of the gauge parameters
$\epsilon_i$ we make the following redefinitions
\begin{eqnarray}
  \label{redef}
   A^\alpha_1 & = & \tilde A^\alpha_1 + b^\alpha V^2 \; , \qquad 
   A^\alpha_2 = \tilde A^\alpha_2 - b^\alpha V^1 \; , \nn \\
   B^\alpha & = & \tilde B^\alpha - \tilde A_i \wedge V^i + b^\alpha
   V^1 \wedge V^2 \; .
\end{eqnarray}
Note that these definitions for the four-dimensional fields can be
obtained directly by expanding the six-dimensional fields $\hat
B^\alpha$ in the basis $\eta^i$. With these definitions we otain
\begin{eqnarray}
  \label{HF1}
  H^\alpha & = & d \tilde B^\alpha  + \tilde A_i^\alpha\wedge dV^i +
  M_i^\alpha{}_\beta \tilde B^\beta \wedge V^i \; ; \nn \\
  F^\alpha_1 & = & \tilde F^\alpha_1 + b^\alpha dV^2 \; , \quad 
  F^\alpha_2  =  \tilde F_2 ^\alpha  - b^\alpha dV^1 \\ 
  D b^\alpha & = & d b^\alpha - M_1^\alpha{}_\beta \tilde A^\beta_2 +
  M_2^\alpha{}_\beta \tilde A^\beta_1 + M_i^\alpha{}_\beta b^\beta V^i  \nn
\end{eqnarray}
where $\tilde F_1^\alpha$ and $\tilde F_2^\alpha$ are defined as
\begin{equation}
  \label{F12}
  \tilde F_i^\alpha = d \tilde A^\alpha_i - M_j^\alpha{}_\beta \tilde
  A^\beta_i \wedge V^j - M_i^\alpha{}_\beta  \tilde B^\beta \; .
\end{equation}
For these new fields the gauge transformations read
\begin{eqnarray}
  \label{gaugetrsf3}
  \delta V^i & = & d \epsilon^i \; ;\\
  \delta b^\alpha & = & -M_i^\alpha{}_\beta b^\beta \epsilon^i \; ; \qquad
  \delta \tilde A^\alpha_i = - M_j^\alpha{}_\beta \tilde A_i^\beta
  \epsilon^j \; ;\qquad 
  \delta \tilde B^\alpha = - M_i^\alpha{}_\beta \tilde B^\beta \epsilon^i \; .
\end{eqnarray}
The field strengths transform covariantly
\begin{eqnarray}
  \label{fsgtr}
  \delta Db^\alpha = -M_i^\alpha{}_\beta Db^\beta \epsilon^i \; ; \quad
  \delta \tilde F_i^\alpha = - M_j^\alpha{}_\beta \tilde F_i^\beta
  \epsilon^j \; ; \quad \delta \tilde H^\alpha = -M_i^\alpha{}_\beta
  \tilde H^\beta \epsilon^i \; ,
\end{eqnarray}
and in order to show this we needed again the fact that the matrices $M_1$
and $M_2$ commute.

\subsection{Imposing the self-duality conditions}

The final step, in order to obtain the final four-dimensional action
would be to elliminate the doubled degrees of 
freedom. Note that we shall use the definitions \eqref{fsdef} and
ignore for the moment \eqref{redef}. 
Let us see first which are the fields we would
like to keep in the final theory. Regarding the gauge fields it should
not be important whether we keep $A_1$ or $A_2$ as they appear in a
rather symmetric fashion. Suppose we keep $A_1$. There is no reason
apriori to 
consider some linear combination of $A_1$ and $A_2$. 
Recall that in general flux compactifications the kinetic terms are
not modified compared to the usual massless compactifications. This is
the case for the gauge fields $A_1$ or $A_2$. A linear combination of
the gauge fields would make sense if the coefficients are related to
the twist matrices so that other parts of the action may be
simplified. However such a field redefinition would introduce the
twist parameters in the kinetic terms and would put the action in a
non-standard form. Shortly we shall motivate on other grounds a
twist-dependent redefinition of the gauge fields. 

Now let us consider the fields $B^\ax$ and $b^\ax$. From the form of
the field strengths \eqref{fsdef} it is be clear that the tensor
fields are massive due to the Stuckelberg couplings to the vector
fields \cite{LM1,DASV,SV}. This means that we have to keep the 
tensor fields and elliminate the scalars $b^\ax$. Trying to remove the
tensor fields from the spectrum would result into scalars which are
both electrically and magnetically charged, as it can be seen from their
covariant derivative. 

The strategy, in order to write the action in terms of $A^\ax_1$ and
$B^\ax$, is to add suitable total derivative terms to the action such
that the variation with respect to $F^\ax_2$ and
$Db^\ax$ reproduces the self-duality constraints. Ellimination of the
fields $F^\ax_2$ and $Db^\ax$ from the action would then give the
desired result. The terms we will add are of the form $\rho_{\ax \bx}
dB^\ax \wedge db^\beta$ and $\rho_{\ax\bx} dA_1^\ax \wedge dA_2^\bx$
and in order to obtain the self-duality relations we would like to
express these terms in terms of the field-strengths $F_i^\ax$, $H^\ax$
and $Db^\ax$. It is straightforward to check that
\begin{eqnarray}
  S_d =
  \rho_{\ax\bx} H^\ax \wedge Db^\bx + \rho_{\ax\bx} F^\ax_1 \wedge
  F^\bx_2 -2 \rho_{\ax \bx} M_2^\bx{}_\cx d A_1^\ax \wedge B^\cx -
  \rho_{\ax \bx} M_1^\ax{}_\dx M_2^\bx{}_\cx B^\dx \wedge B^\cx 
  \nonumber \\
  = \rho_{\ax \bx} dB^\ax \wedge db^\beta + \rho_{\ax\bx} dA_1^\ax
  \wedge dA_2^\bx - \rho_{\ax\bx}M_2^\bx{}_\cx d(A_1^\ax \wedge B^\cx)
  + \rho_{\ax \bx} M_1^\bx{}_\cx d(A_2^\ax \wedge B^\cx) \; ,  \nonumber
\end{eqnarray}
and therefore $S_d$ is a total derivative. Let us now consider 
\begin{equation}
  S_{Total} = S_T - \tfrac12 S_d \; .
\end{equation}
Taking variations of this total action with respect to $Db^\ax$ and
$F_2^\ax$ reproduces the self-duality constraints \eqref{4dsd}.
Replacing these constraints into the total action, we see that 
$S_T$ identically vanishes as it should have already
happened in six dimensions had we imposed the self-duality constraints
\eqref{SD} in the action \eqref{6dact}. Therefore, the
only piece we have to deal with is $S_d$. This becomes
\begin{equation}
  \label{finact}
  \begin{aligned}
      S_d = &- \sqrt G g_{\ax \bx} H^\ax \wedge * H^\bx -
      \frac1{G^{22}} g_{\ax \bx} F_1^\ax \wedge * F_1 ^\bx +
      \frac{G^{12}}{G^{22}} \rho_{\ax \bx} F_1^\ax \wedge F_1^\bx \\ 
      & -2 \rho_{\ax \bx} M_2^\bx{}_\cx d A_1^\ax \wedge B^\cx - 
      \rho_{\ax \bx} M_1^\ax{}_\dx M_2^\bx{}_\cx B^\dx \wedge B^\cx \; .
  \end{aligned}
\end{equation}
As anticipated, we end up with a theory for tensor fields
which acquire a mass via the Stuckelberg mechanism
\begin{equation}
  \label{gtr}
  \delta B^\ax = d \Lambda^\ax \; ; \qquad \delta A_1^\ax = -
  M_1^\ax{}_\bx \Lambda^\bx \; ,
\end{equation}
where $\Lambda^\ax$ are 1-form gauge parameters. However in trying to
replace the self-duality conditions \eqref{4dsd} in the field
strengths \eqref{fsdef} we obtain cyclic definitions for $H^\ax$.
This situation resembles somewhat the results in \cite{andrianopoli}
where it was found that in $N=2$ supergravity coupled to vector-tensor
multiplets the Bianchi identities require to introduce magnetic dual
degrees of freedom.  We may try to fix this problem by implementing
the redefinitions \eqref{redef} and \eqref{F12}. However in this way
the fields $b^\ax$ will appear in the action without derivative and
their replacement using teh self-duality conditions \eqref{4dsd} will
no longer be possible. In fact, this can be seen in a more clear way
by considering a completely massless compactification where both twist
matrices $M_1$ and $M_2$ vanish. Even in this case the ellimination of
the scalars $b^\ax$ in the favour of the tensor fields $B^\ax$ can not
be done consistently. The gauge fields $A_1$ still have to be
redefined according to \eqref{redef}. In this way, the scalars $b^\ax$
appear in the gauge coupling matrix (not only in the generalised
$\theta$ angles) which proves they are not axions and so we can not
expect to be able to dualise them to tensor fields in the usual way.
Therefore, we can argue that we have to keep the scalar fields in the
resulting theory. However, as we have explained before, in the case
that both twist matrices are non-vanishing, the tensor fields are
massive and we should rather keep them and not the scalar fields. The
way out from this puzzle is to find a different symplectic gauge for
the gauge fields where the tensor fields are not explicitely massive
and where one can safely remove them from the spectrum.  One obvious
choice would be to consider as the electric gauge fields the
combination which appears in the covariant derivative $Db^\ax$ in
\eqref{fsdef}. Let us define
\begin{equation}
  A_-^\ax = M_1^\ax{}_\bx A_2^\bx - M_2^\ax{}_\bx A_1^\bx \; .
\end{equation}
Note that, in the corresponding field strength, the tensor fields appear
as $[M_1, M_2]^\ax{}_\bx B^\bx$ which vanishes due to the fact that the matrices
$M_1$ and $M_2$ commute. Therefore, $A_-^\ax$ are suitable candidates
for electric gauge fields. This analysis can be carried out in full
generality, but in order to point out the main features we shall choose
a particular case, $M_1= M_2 = M$, which is technically less
involved. In this particular case we can redefine the gauge fields as
\begin{equation}
  \label{A+-}
  A_\pm = A_1 \pm A_2 \; .
\end{equation}
The field strengths \eqref{fsdef} become
\begin{eqnarray}
  H^\alpha & = & dB^\ax + F_+^\ax \wedge V^+  + F^- \wedge V^- +
  Db^\ax \wedge V^+ \wedge V^- \; ; \nn \\
  F_+^\alpha & = & dA_+^\ax + 2 M^\ax{}_\bx B^\bx + Db^\ax \wedge V^-
  \; ; \nn \\
  F_-^\alpha & = & dA_-^\ax - Db^\ax \wedge V^+
  \; ;\\
  Db^\ax & = & d b^\ax+ M^\ax{}_\bx A_-^\bx\; , \quad D v^\ax = d
  v^\ax + M^\ax{}_\bx v^\bx V^+
  \; ; \nn
\end{eqnarray}
where $V^\pm = V^1 \pm V^2$. The action has precisely the same form as
\eqref{4dact} where now the indices $i, j$ are understood to take the
values $\pm$ and the metric is given by
\begin{equation}
  G_{ij} = \frac14 \left(
  \begin{array}{cc}
    G^{11} + 2 G^{12}+ G^{22} & G^{11} - G^{22} \\
    G^{11}-G^{22} & G^{11} - 2 G^{12}+ G^{22}
  \end{array}
  \right)
\end{equation}

The field
strengths above suggest that it should be possible to keep the gauge
fields $A_-^\ax$ together with the scalars $b^\ax$ and elliminate
$H^\ax$ and $F_+^\ax$. As before we add a total derivative
\begin{eqnarray}
  S_d & = & \rho_{\ax \bx} H^\ax \wedge Db^\bx - \tfrac12 \rho_{\ax
    \bx} F_+^\ax \wedge F_-^\bx - \rho_{\ax\bx} dA_-^\ax \wedge V^-
  \wedge Db^\bx \\
  & & = \rho_{\ax\bx} \big( dB^\ax \wedge db^\bx - \tfrac12 dA_+
  \wedge dA_- \big) - \rho_{\ax\bx} M^\bx{}_cx d(B^\ax \wedge A_-^\cx) 
  \nn
\end{eqnarray}
and one can again check that the self-duality conditions written in
the $\pm$ basis can be obtained by taking variations of the total
action with respect to $F_+^\ax$ and $H^\ax$. Replacing the
self-duality conditions in $S_d$ we obtain
\begin{eqnarray}
  \label{SN2}
  S_d & = & - \frac{1}{\sqrt G} g_{\ax\bx} Db^\ax \wedge * Db^\bx 
  -\frac1{G^{++} \sqrt G} g_{\ax\bx} F_-^\ax \wedge *F_-^\bx \nn \\
  & & + \tfrac12 \rho_{\ax\bx} \frac{G^{+-}}{G^{++}} F_-^\ax \wedge F_-^\bx
  - \rho_{\ax\bx} dA_-^\ax \wedge V^- \wedge Db^\bx \; .
\end{eqnarray}

With a little bit of effort, the action above can be put in the
standard $N=2$ gauged supergravity form \cite{N2}. We shall not do it
explicitely, but we shall just describe the steps which are rather
standard. First of all one redefines the gauge fields as
$A_-^\ax \to A_-^\ax + b^\ax V^+$. Then the last term in the action
above can be integrated by parts
\begin{equation}
  \rho_{\ax\bx} dA_-^\ax \wedge V^- \wedge Db^\bx = \rho_{\ax\bx}
  b^\bx d A_-^\ax \wedge dV^- + \tfrac12 M_{\ax\bx} A_-^\ax \wedge
  A_-^\bx \wedge dV^- + \mathrm{total~derivative}\; .
\end{equation}
One can therefore dualize the gauge field $V^-$ to its magnetic dual
$\tilde V^-$ whose field strength will be of the form $d \tilde V^- +
\tfrac12 M_{\ax\bx} A_-^\ax \wedge A_-^\bx$. Finally we have to go to
the Einstein frame in the action \eqref{SR} and redefine the fields
$v^\ax$ as
\begin{equation}
  v^\ax = \frac1{\sqrt G} \tilde v^\ax \; .
\end{equation}
This effectively means that one of the $T^2$ moduli, namely $\sqrt G$
becomes part of the scalars $\tilde v^\ax$ which will no longer be
constrained. We can now write the combination $t^\ax = b^\ax + i
v^\ax$ which will have the kinetic term $g_{\ax\bx} D t^\ax \wedge * D
t^\bx$ where the covariant derivatives are given by
\begin{equation}
  D t^\ax = d t^\ax + M^\ax{}_\bx t^\bx V^+ + M^\ax{}_\bx A^\bx_- \; .
\end{equation}

To conclude this section we mention that in the action \eqref{SN2} the
kinetic terms for the gauge fields depend explicitely on the choice of
fluxes $M_i$. This may not be completely clear due to the choice we
made -- $M_1 = M_2$ -- in writing the action \eqref{SN2}. It is clear
however that in the general case this action will not look so simple
and moreover the twist matrices will appear explicitely in the kinetic
terms for the gauge fields.

\subsection{Conclusions}

In this note we derived part of the action which comes from the
compactification of F-theory on certain manifolds with $SU(3)$
structure. We argued that the compactification can be reduced to a
Scherk-Schwarz compactification of six-dimensional supergravity. The
direct result is a $N=2$ gauged supergravity coupled to vector-tensor
multiplets and we have seen that in such a case one can not completely
remove the magnetic dual degrees of freedom from the action which is
in agreement with the results found in \cite{andrianopoli}. In a
suitable chosen basis for the gauge fields, the magnetic dual degrees
of freedom can be decoupled and we end up with ordinary $N=2$ gauged
supergravity. However, from a physical perspective, The first
formulation in terms of vector-tensor multiplets might be more sensible as
the usual supergravity quantities (gauge coupling functions in
particular) are just given in terms of the geometric data of the
compactification manifold as it is the case in massless
compactifications which is not the case with the action \eqref{SN2}.
The same point of view may be sustained from the
string duality perspective as the dualities are first established at
the massless level and only afterwards are deformed to accommodate
fluxes. On the other hand we are not aware of any string
compactification where vector-tensor multiplets appear non-trivially
and therefore therefore the analysis in this paper opens the quest for
other compactifications which involve vector-tensor multiplets.

{\textbf{Acknowledgments}}
This work was supported in part by the National University Research
Council CNCSIS-UEFISCSU, project number PN II-RU 77/04.08.2010 and PN
II-ID 464/15.01.2009 and in part by project ''Nucleu'' PN 09 37 01 02
and PN 09 37 01 06.


\begin{thebibliography}{99}

\bibitem{ABLM}
  O.~Aharony, M.~Berkooz, J.~Louis, A.~Micu,
  \emph{Non-Abelian structures in compactifications of M-theory on
    seven-manifolds with SU(3) structure},
  JHEP {\bf 0809 } (2008)  108.
  [arXiv:0806.1051 [hep-th]].

\bibitem{AM1}
  A.~Micu,
  \emph{Heterotic type IIA duality with fluxes - towards the complete story},
  JHEP {\bf 1010 } (2010)  059.
  [arXiv:1009.2357 [hep-th]].

\bibitem{RES}
  R.~A.~Reid-Edwards, B.~Spanjaard,
 \emph{N=4 Gauged Supergravity from Duality-Twist Compactifications
    of String Theory}, 
  JHEP {\bf 0812 } (2008)  052.
  [arXiv:0810.4699 [hep-th]].

\bibitem{STW} 
  J.~Shelton, W.~Taylor and B.~Wecht,
  ``Nongeometric flux compactifications,''
  JHEP\ {\bf 0510}, 085  (2005)
  [hep-th/0508133].


\bibitem{SS}
  J.~Scherk and J.~H.~Schwarz,
  ``How to Get Masses from Extra Dimensions,''
  Nucl.\ Phys.\ B\ {\bf 153} (1979) 61.

\bibitem{romans}
  L.~J.~Romans,
  ``Selfduality For Interacting Fields: Covariant Field Equations For
  Six-dimensional Chiral Supergravities,'' 
  Nucl.\ Phys.\ B\ {\bf 276} (1986) 71.

\bibitem{thomas}
  F.~Bonetti, T.~W.~Grimm,
  \emph{Six-dimensional (1,0) effective action of F-theory via
    M-theory on Calabi-Yau threefolds} 
arXiv:1112.1082 [hep-th]

\bibitem{MS}
  N.~Marcus and J.~H.~Schwarz,
  ``Field Theories That Have No Manifestly Lorentz Invariant Formulation,''
  Phys.\ Lett.\ B\ {\bf 115} (1982) 111.

\bibitem{dallAgata}
  G.~Dall'Agata,
  ``Type IIB supergravity compactified on a Calabi-Yau manifold with H fluxes,''
  JHEP\ {\bf 0111} (2001) 005
  [hep-th/0107264].


\bibitem{LM1}
  J.~Louis, A.~Micu,
  \emph{Type 2 theories compactified on Calabi-Yau threefolds in the
    presence of background fluxes}, 
  Nucl.\ Phys.\  {\bf B635 } (2002)  395-431.
  [hep-th/0202168].


\bibitem{DASV}
  G.~Dall'Agata, R.~D'Auria, L.~Sommovigo and S.~Vaula,
  ``D = 4, N=2 gauged supergravity in the presence of tensor multiplets,''
  Nucl.\ Phys.\ B\ {\bf 682} (2004) 243
  [hep-th/0312210].

\bibitem{SV}
  L.~Sommovigo, S.~Vaula,
  \emph{D=4, N=2 supergravity with Abelian electric and magnetic charge},
  Phys.\ Lett.\  {\bf B602 } (2004)  130-136.
  [hep-th/0407205].

\bibitem{andrianopoli}
  L.~Andrianopoli, R.~D'Auria, L.~Sommovigo, M.~Trigiante,
  \emph{D=4, N=2 Gauged Supergravity coupled to Vector-Tensor Multiplets},
  Nucl.\ Phys.\  {\bf B851 } (2011)  1-29.
  [arXiv:1103.4813 [hep-th]].

\bibitem{N2}
  L.~Andrianopoli, M.~Bertolini, A.~Ceresole, R.~D'Auria, S.~Ferrara, P.~Fre and T.~Magri,
  ``N = 2 supergravity and N = 2 super Yang-Mills theory on general scalar
  manifolds: Symplectic covariance, gaugings and the momentum map,''
  J.\ Geom.\ Phys.\  {\bf 23} (1997) 111
  [arXiv:hep-th/9605032].


\end{thebibliography}
\end{document}